\documentstyle[prl,aps,twocolumn,psfig,epsf]{revtex}

\begin{document}
\title{Long-range proximity effects in superconductor-ferromagnet structures }
\author{F. S. Bergeret $^{1 }$, A. F. Volkov$^{1,2}$ and K. B. Efetov$^{1,3}$}
\address{$^{(1)}$Theoretische Physik III,\\
Ruhr-Universit\"{a}t Bochum, D-44780 Bochum, Germany\\
$^{(2)}$Institute of Radioengineering and Electronics of the Russian Academy%
\\
of Sciences, 103907 Moscow, Russia \\
$^{(3)}$L.D. Landau Institute for Theoretical Physics, 117940 Moscow, Russia }
\maketitle

\begin{abstract}
We analyze the proximity effect in a superconductor/ferromagnet (S/F)
structure with a local inhomogeneity of the magnetization in the ferromagnet
near the S/F interface. We demonstrate that not only the singlet but also
the triplet component of the superconducting condensate is induced in the
ferromagnet due to the proximity effect. Although the singlet component of
the condensate penetrates into the ferromagnet over a short length $\xi _{h}=%
\sqrt{D/h}$ ($h$ is the exchange field in the ferromagnet and $D$ the
diffusion coefficient), the triplet component, being of the order of the
singlet one at the S/F interface, penetrates over a long length $\sqrt{%
D/\epsilon }$ ($\epsilon$ is the energy). This long-range penetration leads
to a significant increase of the ferromagnet conductance below the
superconducting critical temperature $T_{c}$.
\end{abstract}

In recent experiments on S/F structures a considerable increase of the
conductance below the superconducting critical temperature $T_{c}$ was
observed \cite{Petrashov1,Pannetier,Chandra}. Although in a recent work \cite
{Belzig} it was suggested that such an increase may be due to scattering at
the S/F interface, a careful measurement of the conductance demonstrated
that the entire change of the conductance was due to an increase of the
conductivity of the ferromagnet \cite{Petrashov1,Pannetier}.

Such an increase would not be a great surprise if instead of the ferromagnet
one had a normal metal N. It is well known (see for review \cite{Been,Lamb})
that in S/N structures proximity effects can lead to a considerable increase
of the conductance of the N wire provided its length does not exceed the
phase breaking length $L_\varphi$. However in  a S/F structure, if the
superconducting pairing is singlet, the proximity effect is negligible at
distances exceeding a much shorter length $\sim \xi_h$. This reduction of
the proximity effect due to the exchange field $h$ of the ferromagnet is
clear from the picture of Cooper pairs consisting of electrons with opposite
spins. The proximity effect is not considerably affected by the exchange
energy only if the latter is small $h<T_{c}$. As concerns such strong
ferromagnets as $Fe $ or $Co$ used in the experiments \cite
{Petrashov1,Pannetier}, whose exchange energy $h$ is by several orders of
magnitude larger than $T_{c}$, a singlet pairing is impossible due to the
strong  difference in the energy dispersions for the two spin bands. At the
same time, an arbitrary exchange field cannot destroy a triplet
superconducting pairing because the spins of the electrons forming Cooper
pairs are already parallel. A possible role of the triplet component in
transport properties of S/F structures has been noticed in Refs. \cite{Falko,Spivak},  where the triplet component arose only  as a result of
mesoscopic fluctuations. However, in both cases the corrections to the
conductance are much smaller than the observed ones.

In this paper, we suggest a much more robust mechanism of formation of the
triplet pairing in S/F structures, which is due to a local inhomogeneity of
the magnetization $M$ in the vicinity of the S/F interface. We show that the
inhomogeneity generates a triplet component of the superconducting order
parameter with an amplitude comparable with that of the singlet pairing. The
penetration length of the triplet component into the ferromagnet is equal to 
$\xi _{\varepsilon }=\sqrt{D/\varepsilon }$, where the energy $\varepsilon $
is of the order of temperature $T$ or the Thouless energy $E_{T}=D/L^{2},$ $L
$ is the sample size. The length $\xi _{\varepsilon }$ is of the same order
as that for the penetration of the superconducting pairs into a normal metal
and therefore the increase of the conductance due to the proximity effect
can be comparable with that in an S/N structure.

\begin{figure}
\epsfysize = 4.5cm

\centerline{\epsfbox{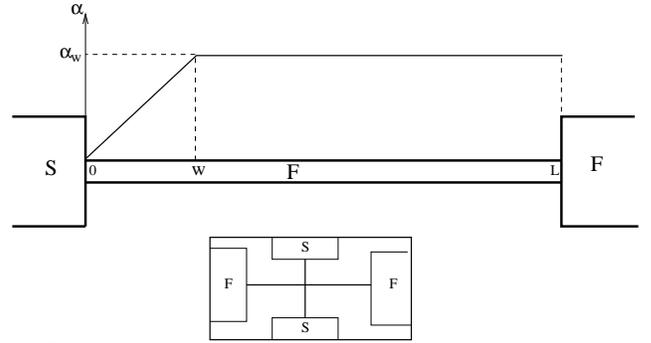
 }}
\caption{Schematic view of the structure under consideration. In the inset is
 shown the structure, for which we calculate the conductance variation: two
 ferromagnetic wires connected to two ferromagnetic and two superconducting reservoirs. \label{Fig.1}}
\end{figure}


We consider a structure shown in Fig.1 and assume that the magnetization
orientation varies linearly from $\alpha =0$ at $x=0$ to $\alpha_{w}= Qw$ at 
$x=w$. Here $\alpha$ is the angle between $M$ and the $z$-axis. The case $%
Qw=\pi $ corresponds to a domain wall with thickness $w$ located at the S/F
interface, while the model with the homogeneous magnetization is recovered
by putting $Q=0$. This variation of $M$ may also be brought about by an
external magnetic field (see \cite{Waintal} and references therein). Of
course, the variation of the magnetization considered is the simplest model
of what may happen at the interface in reality and we use it for simplicity.
We consider the diffusive limit corresponding to a short mean free path. In
this limit one may describe the S/F structure using the Usadel equation \cite
{Usadel}. The proximity effect in a S/F structure with an uniform
magnetization $M$ was analyzed in \cite{Demler}. For the system with a
non-homogeneous magnetization one should use a generalized form of this
equation containing spin variables \cite{Berger}. The equation is non-linear
and contains normal $\check{g}$ and anomalous $\check{f}$ quasiclassical
Green's functions. These functions are $4\times 4$ matrices in the Nambu$%
\otimes $spin space.

Assuming that the anomalous condensate function $f$ is small, one can
linearize the Usadel equation. This can be done if the transmission
coefficient through the S/F interface is small due to a strong mismatch of
the Fermi surfaces. Moreover, the order parameter in the superconductor can
be strongly suppressed near the S/F interface if the transparency is high.
In both cases one may assume a weak proximity effect, which presumably
corresponds to the experiments. We write the Usadel equation for the matrix
element (12) of $\check{f}$ in the Nambu space. Then for the retarded matrix
(in the spin space) Green's function $\hat{f}^R$ we obtain (the index $R$ is
dropped)

\begin{equation}
-iD\partial _{{\bf r}}^{2}\hat{f}+2\epsilon \hat{f}-2\Delta \hat{\sigma}%
_{3}+\left( \hat{f}\hat{V}^{\ast }+\hat{V}\hat{f}\right) =0  \label{a2}
\end{equation}
Here $\epsilon $ is the energy, $\Delta $ is the superconducting order
parameter, which vanishes in the ferromagnet; $\hat{\sigma}_{i}$ are the
Pauli matrices in the spin space, and the matrix $\hat{V}$ is defined as $%
\hat{V}=h\left( \hat{\sigma}_{3}\cos \alpha +\hat{\sigma}_{2}\sin \alpha
\right) $, where $\alpha $ varies with $x$ as shown in Fig.1. This matrix
describes the interaction between the exchange field and spins of the
conduction electrons and vanishes in the superconductor. Eq. (\ref{a2}) could be written for temperature anomalous Green
functions $\hat{f}^{M}$  at Matsubara frequencies $\omega $ , by
replacing $\epsilon \rightarrow i\left| \omega \right| $ \ and
multiplying the last term by ${\rm sgn}\left( \omega \right) $ \cite{Berger}. Eq. (\ref{a2}) is supplemented by the boundary conditions at the interface
that can also be linearized \cite{Zaitsev}. Assuming that there are no spin-flip processes at
the S/F interface, we have 
\begin{equation}
\left. \partial _{x}\hat{f}\right| _{x=0}=\left( \rho /R_{b}\right) \hat{f}%
_{S}\;,  \label{bound-cond}
\end{equation}
where $\rho $ is the resistivity of the ferromagnet, $R_{b}$ is the S/F
interface resistance per unit area in the normal state, and $f_{S}=\hat{%
\sigma}_{3}\Delta /\sqrt{\epsilon ^{2}-\Delta ^{2}}$. The solution of Eq. (\ref{a2}) is trivial in the superconductor but needs
some care in the ferromagnet. In the region $0<x<w$ the solution $\hat{f}$
can be sought in the form $\hat{f}=\hat{U}\left( x\right) \hat{f}_{n}\hat{U}%
\left( x\right)$, where $\hat{U}\left( x\right) =\hat{\sigma}_{0}\cos \left(
Qx/2\right) +i\hat{\sigma}_{1}\sin \left( Qx/2\right)$. Substituting this
expression into Eq. (\ref{a2}) and assuming that the solution depends on the
coordinate $x$ only we obtain the following equation for $\hat{f}_{n}$ 
\begin{eqnarray}
-iD\partial_{xx}^2\hat{f}_n\!\!+\!\!i\left(DQ^2/2\right)\left(\hat{f}_n+\hat{%
\sigma}_1 \hat{f}_n\hat{\sigma}_1\right)\!\!+\!\!DQ\left\{\partial_{x}\hat{f}%
_n,\hat{\sigma}_1\right\}  \nonumber \\
+2\epsilon\hat{f}_n+h\left\{\hat{\sigma}_3,\hat{f}_n\right\}\!=0  \label{a5}
\end{eqnarray}
Here $\{...\}$ is the anticommutator. In the region $x>w$,  $\hat{f}_{n}$
satisfies Eq. (\ref{a5}) with  $Q=0$.

We see from Eq. (\ref{a5}) that the singlet component, commuting with $\hat{%
\sigma}_3$, and triplet component, anticommuting with $\hat{\sigma}_3$, are
mixed by the rotating exchange field $h$. In the region $x>w$ the triplet
and the singlet components decouple and their amplitudes should be found by
matching the solutions at $x=w$. One should also use the boundary condition,
Eq. (\ref{bound-cond}), and match the solutions in the ferromagnet and
superconductor. It is clear that the singlet and triplet components of the
anomalous function $\hat{f}_{n}$ inevitably coexist in the ferromagnet. This
fact is also known for the case of magnetic superconductors with $Q\neq 0$ 
\cite{Bul}. In the region $x>w$ the singlet part decays sharply but the
triplet one survives over long distances. We are able to confirm these
statements solving Eq. (\ref{a5}) with the boundary condition Eq. (\ref
{bound-cond}). In the case of a homogeneous magnetization ($Q=0$) the
triplet pairing cannot be induced, which follows immediately from Eq. (\ref
{bound-cond}) connecting separately the singlet and triplet components at
the opposite sides of the interface and Eq. (\ref{a5}).

Eq. (\ref{a5}) can be solved exactly. The solution $\hat{f}_{n}$ can be
written in the form 
\begin{equation}
\hat{f}_{n}= \hat{\sigma}_0A\left( x\right) +\hat{\sigma}_{3}B\left(
x\right)+i\hat{\sigma}_{1}C\left( x\right)  \label{2}
\end{equation}

The function $C\left( x\right) $ in Eq. (\ref{2}) is the amplitude of the
triplet pairing, whereas the first and the second term describe the singlet
one. Substituting Eq. (\ref{2}) into Eq. (\ref{a5}) we obtain a system of
three equations for the functions $A $, $B$
and $C $, which can be sought in the form 
\begin{equation}
A\left( x\right) =\sum_{i=1}^{3}\left( A_{i}\exp \left( -\kappa _{i}x\right)
+\bar{A}_{i}\exp \left( \kappa _{i}x\right) \right)  \label{3}
\end{equation}
The functions $B(x)$ and $C(x)$ can be written in a similar way. The
eigenvalues $\kappa _{i}$ obey the algebraic equations 

\begin{eqnarray}
\left( \kappa ^{2}-\kappa _{\epsilon }^{2}-Q^{2}\right) C-2\left( Q\kappa
\right) A &=&0  \nonumber \\
\left( \kappa ^{2}-\kappa _{\epsilon }^{2}\right) B-\kappa _{h}^{2}A &=&0
\label{4} \\
\left( \kappa ^{2}-\kappa _{\epsilon }^{2}-Q^{2}\right) A-\kappa
_{h}^{2}B+2\left( Q\kappa \right) C &=&0\;,  \nonumber
\end{eqnarray}
where $\kappa _{\epsilon }^{2}=-2i\epsilon /D$ and $\kappa _{h}^{2}=-2ih/D$
(indices $i$ were dropped).
The eigenvalues $\kappa $ are the values at which the determinant of Eqs. (%
\ref{4}) turns to zero. From the first equation of Eqs. (\ref{4}) we see
that in the homogeneous case ($Q=0$) the triplet component has a
characteristic penetration length $\sim \kappa _{\epsilon }^{-1}$, but we
see from Eq. (\ref{bound-cond}) that its amplitude is zero. If $Q\not=0$,
the triplet component $C$ is coupled to the singlet component ($A$, $B$)
induced in the ferromagnet according to the boundary condition Eq. (\ref
{bound-cond}) (proximity effect). If the width $w$ is small, the triplet
component changes only a little in the region $(0,w)$ and spreads over a
large distance of the order $\left| \kappa _{\epsilon }^{-1}\right| $ in the
region $(0,L)$. In the case of a strong exchange field $h$, $\xi _{h}$ is
very short ($\xi _{h}\ll w,\xi _{T}$), the singlet component decays very
fast over the length $\xi _{h}$, and its slowly varying part turns out to be
small. In this case the first two eigenvalues $\kappa _{1,2}\approx(1\pm
i)/\xi _{h}$ can be used everywhere in the ferromagnet $\left( 0<x<L\right) $%
, where $L$ is the length of the ferromagnet. As concerns the third
eigenvalues, we obtain $\kappa _{3}=\sqrt{\kappa _{\epsilon }^{2}+Q^{2}}$ in
the interval $(0,w)$, and $\kappa _{3}=\kappa _{\epsilon }$ in the interval $%
(w,L)$. The amplitude $B_{3}$ of the slowly varying part of the singlet
component is equal to $B_{3}=2\left( Q\kappa _{3}/\kappa _{h}^{2}\right)
C_{3}\ll C_{3}$.

All the amplitudes should be chosen to satisfy the boundary conditions at $%
x=0$ (Eq. (\ref{bound-cond})) and zero boundary condition at $x=L$.  For
the triplet component we obtain (we restore the indices R(A)) 
\begin{eqnarray}
\lefteqn{C^{R(A)}(x)=\mp i\left\{ QB(0)\sinh \left( \kappa _{\epsilon
}(L-x)\right) .\right. }  \nonumber \\
&\!\!\!\!\left. \left[ \kappa _{\epsilon }\cosh \Theta _{\epsilon }\cosh
\Theta _{3}+\kappa _{3}\sinh \Theta _{\epsilon }\sinh \Theta _{3}\right]
^{-1}\!\right\} ^{R(A)}&  \label{5}
\end{eqnarray}
where $w<x<L$, $B^{R(A)}(0)\!\!\!=\!\!\left( \rho \xi _{h}/2R_{b}\right)
f_{S}^{R(A)}$ is the amplitude of the singlet component at the S/F
interface, $\Theta _{\epsilon }\!\!\!=\!\!\kappa _{\epsilon }L$, $\Theta
_{3}\!\!\!=\!\!\kappa _{3}w$, and $\kappa _{\epsilon }^{R(A)}\!\!\!=\!\!%
\sqrt{\mp 2i\epsilon /D}$. One can see that the difference $C^{R}-C^{A}$ is
an even function of $\epsilon $. This is a direct consequence of the fact
that $C^{R}-C^{A}$ is proportional to the Fourier transform of the
correlator $K(t)\!\!\!=\!\!\left\langle \psi _{\uparrow }(t)\psi _{\uparrow
}(0)+\psi _{\uparrow }(0)\psi _{\uparrow }(t)\right\rangle $, which is even
in time. In the Matsubara representation, $C^{R(A)}$ in Eq. (\ref{5}) should
be replaced by $C_{\omega }$ with ${\rm sgn}\omega $ instead of $(\mp )$ and 
$\kappa _{\omega }\!\!\!=\!\!\sqrt{|\omega |/D}$, $f_{S}(\omega
)\!\!\!=\!\!\Delta /\sqrt{\omega ^{2}+\Delta ^{2}}$ instead of $\kappa
_{\epsilon }^{R(A)}$, $f_{S}^{R(A)}$ respectively. Thus, $C_{\omega }$
corresponding to the temperature correlator ${\cal K}\left( \tau \right)
=-<T_{\tau }\psi _{\uparrow }\left( 0\right) \psi _{\uparrow }\left( \tau
\right) >$ is an odd function of $\omega $ and the sum over all $%
\omega $ is zero in accordance with $K(0)\!\!={\cal K}\left( 0\right) =\!0$.

It is clear from Eq. (\ref{5}), that the triplet component is of the same
order of magnitude as the singlet one at the interface. Indeed, for the case 
$w\ll L$ we obtain from Eq. (\ref{5}) $\left| C(0)\right| \sim B(0)/\sinh
\alpha _{w}$, where $\alpha _{w}=Qw$ is the angle characterizing the
rotation of the magnetization. Therefore if the angle $\alpha _{w}\leq 1$
and the S/F interface transparency is not too small, the singlet and triplet
components are not small. They are of the same order in the vicinity of the
S/F interface, but while the singlet component decays abruptly over a short
distance ($\sim \xi _{h}$), the triplet one varies smoothly along the
ferromagnet, turning to zero at the F reservoir. In Fig.2 we plot the
spatial dependence of the singlet $|B(x)|$ and the triplet $|C(x)|$
components for two different $Q$. One can see that the singlet component
decays abruptly undergoing the well known oscillations \cite{Buzdin2} while
the triplet one decays to zero slowly. This decay in the region $(0,w)$
increases with increasing $Q$.

Thus, we come to a remarkable conclusion: the penetration of the
superconducting condensate into a ferromagnet may be similar to the
penetration into a normal metal. The only difference is that, instead of the
singlet component in the case of the normal metal, the triplet one
penetrates into the ferromagnet. Of course, in order to induce the triplet
component one needs an inhomogeneity of the exchange field at the interface.

The presence of the condensate function (triplet component) in the
ferromagnet can lead to interesting long-range effects. One of them is a
change of the conductance of a ferromagnetic wire in a S/F structure (see
inset in Fig.1) when the temperature is lowered below $T_{c}$. This effect
was observed first in S/N structures and later was successfully explained
(see, e.g. reviews \cite{Been,Lamb}). Now we consider the S/F structure
shown in the inset of Fig.1. The normalized conductance variation $\delta 
\tilde{G}=\left( G-G_{n}\right) /G_{n}$ is given by the expression \cite{VZK}%
: 
\begin{equation}
\delta \tilde{G}=-\frac{1}{32T}{\rm Tr}\int {\rm d}\epsilon F_{V}^{\prime
}\left\langle \left[ \hat{f}^R(x)-\hat{f}^A(x)\right] ^{2}\right\rangle \;.
\label{6}
\end{equation}
Here $G_{n}$ is the  conductance in the normal state, $F_{V}^{\prime }=1/2%
\left[ \cosh ^{-2}((\epsilon +eV)/2T)+\cosh^{-2}((\epsilon -eV)/2T)\right]$,
and $<..>$ denotes the average over the length of the ferromagnetic wire
between the F reservoirs. The function $\hat{f}$ is given by the third term
of Eq. (\ref{2}) with $C^R=-\left(C^A\right)^*$ (we neglect the small
singlet component). Substituting Eqs. (\ref{2}, \ref{5}) into Eq. (\ref{6})
one can determine the temperature dependence $\delta \tilde{G}\left(
T\right) $. Fig.3 shows this dependence. We see that $\delta \tilde{G}$
increases with decreasing temperature and saturates at $T=0$. This monotonic
behaviour of $\delta \tilde{G}$ contrasts with the so called reentrant
behaviour of $\delta \tilde{G}$ in S/N structures \cite{Art,Nazarov} and is
a result of broken time-reversal symmetry of the system under consideration.

Available experimental data are still controversial. It has been established
in a recent experiment \cite{Chandra} that the conductance of the
ferromagnet does not change below $T_{c}$ and all changes in $\delta G$ are
due to changes of the S/F interface resistance $R_b$. However, in other
experiments $R_b$ was negligibly small \cite{Petrashov1}. The mechanism
suggested in our work may explain the long-range effects observed in the
experiments \cite{Petrashov1,Pannetier}. At the same time, the result of the
experiment \cite{Chandra} is not necessarily at odds with our findings. The
inhomogeneity of the magnetic moment at the interface, which is the crucial
ingredient of our theory, is not a phenomenon under control in these
experiments. One can easily imagine that such inhomogeneity existed in the
structures studied in Refs. \cite{Petrashov1,Pannetier} but was absent in
those of Ref. \cite{Chandra}. The magnetic inhomogeneity near the interface
may have different origins. Anyway, a more careful study of the possibility
of a rotating magnetic moment should be performed to clarify this question.

In order to explain the reentrant behaviour of $\delta G(T)$ observed in
Refs. \cite{Petrashov1,Pannetier} one should take into account other
mechanisms, as those analyzed in Refs. \cite{Belzig,Falko,Golubov}. However,
this question is beyond the scope of the present paper.

We note that at the energies $\epsilon $ of the order of Thouless energy $%
\epsilon \sim E_{T}$ the triplet component spreads over the full length $L$
of the ferromagnetic wire (see Fig.2). This long-range effect differs
completely from the proximity effect in a ferromagnet with a uniform
magnetization considered recently in Ref.\cite{Buzdin}. In the latter case
the characteristic wave vector is equal to $\kappa _{1,2}=\sqrt{-2i(\epsilon
\pm h)/D}$ (cf. Eqs. (\ref{4})). It was noted in Ref. \cite{Buzdin} that if $%
\epsilon\rightarrow\pm h$, then $\kappa_{1,2}\rightarrow 0$ and the singlet
component penetrates in the ferromagnet. If the characteristic energies $%
\epsilon_{ch}\sim E_T,T$ are much less than $h$, the penetration length $%
\left| \kappa _{1,2}\right| ^{-1}$ is of the order $\xi _{h}$ and is much
shorter then $\xi _{T}$ or $L$.

\begin{figure}
\epsfysize = 5.2cm

\centerline{\epsfbox{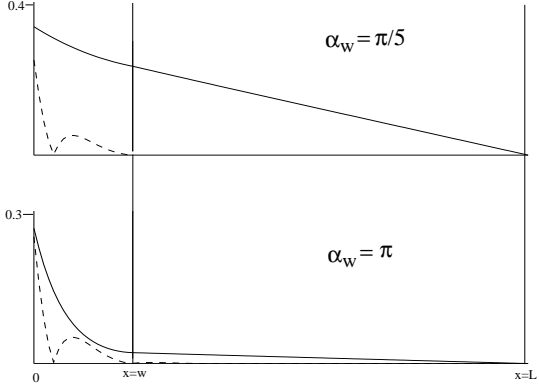
 }}
\vspace{0.2cm}
\caption{Spatial dependence  of the singlet (dashed line) and triplet (solid line)
  components of $\left|\hat{f}\right|$ in the F wire for different values
  of $\alpha_w$. Here $w=L/5$, $\epsilon=E_T$ and $h/E_T=400$. $E_T=D/L^2$ is
  the Thouless energy.}
\label{Fig.2}
\end{figure}


\begin{figure}
\epsfysize = 5.0cm
\centerline{\epsfbox{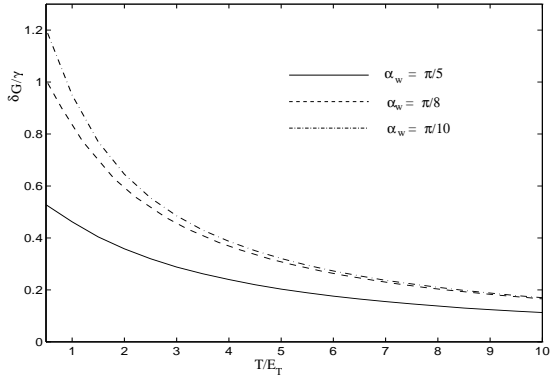
 }}
\vspace{0.2cm}
\caption{The $\protect\delta G(T)$ dependence. Here $\protect\gamma =\protect%
\rho \protect\xi _{h}/R_{b}$. $\Delta /E _{T}\gg 1$ and $%
w/L=0.05$}
\label{Fig.3}
\end{figure}
 It is also interesting to note that a triplet component of the
condensate function with the same symmetry (odd in frequency $\omega $ and
even in momentum $p$) was suggested by Berenziskii \cite{Berez} as a
possible phase in superfluid $^{3}He$ (this, so called ``odd''
superconductivity, was discussed in a subsequent paper \cite{Balatsky}).
Being symmetric in space, this component is not affected by potential
impurities, in contrast to the case analyzed in Ref. \cite{Larkin}, where
the triplet component of the condensate was odd in space. While in $^{3}He$
this hypothetical condensate function is not realized (in $^{3}He$ it is odd
in $p$ but not in frequency), in our system this odd (in $\omega $) triplet
component does exist, although under special conditions described above.

In conclusion, we have shown that in the presence of a local inhomogeneity
of magnetization near the S/F interface, both the singlet and triplet
components of the condensate are created in the ferromagnet due to the
proximity effect. The singlet component penetrates into the ferromagnet over
a short length $\xi _{h}$, whereas the triplet component can spread over the
full mesoscopic length of the ferromagnet. This long-range penetration of
the triplet component should lead to a significant variation of the
ferromagnet conductance below $T_{c}$.

We would like to thank SFB 491 for financial support.

\end{document}